# Aluminium and carbon doped $MgB_2$: band filling, band shift, and anisotropy loss

Sabina Ruiz-Chavarria[a], Gustavo Tavizon[b] and Pablo de la Mora[a1]
[a] Departamento de Física, Facultad de Ciencias,
Universidad Nacional Autónoma de México,
Cd. Universitaria, 04510 Coyoacán, D.F., México
[b] Departamento de Física y Química Teórica, Facultad de Química,
Universidad Nacional Autónoma de México,
Cd. Universitaria, 04510 Coyoacán, D.F., México
[1] e-mail: delamora@servidor.unam.mx

**Abstract**
In this work the effect of Carbon and Aluminium doping on the multiband $MgB_2$ superconductor is analyzed. Using the rigid band and virtual crystal approximations (*RBA* and *VCA*), it was found that the main effect of doping on the band structure is band filling and a relative band-shift. If this band-shift is *eliminated* with an appropriate change of scale, then the *RBA* provides a good description of the band structure as function of doping. With this procedure both the inplane electrical conductivity of the *C*- and *Al*-doped $MgB_2$ and the superconducting critical temperature follow the same curve. The $T_c$ graph approximately follows the $\sigma$-band density of states; the differences between these two can be explained by loss of anisotropy which plays an important role in these systems.

## Introduction

$MgB_2$ has four bands at the Fermi energy ($E_F$), two of them, the $\sigma$-bands, are formed mainly by $B{:}p_x+p_y$ and the other two, the $\pi$-bands, formed mainly by $B{:}p_z$. The superconductivity is due mainly by the $\sigma$-bands; these bands have a strong two dimensional (*2D*) character, while the $\pi$-bands are three dimensional (*3D*).

In the $MgB_2$ superconductor, the $Mg$ atoms can be replaced by $Al$ and by $Sc$ to form $Mg_{1-x}Al_xB_2$ and $Mg_{1-x}Sc_xB_2$; $B$ can be replaced by $C$ to form $MgB_{2-x}C_x$ (these solid solutions will also be denoted as $(Mg,Al)B_2$, $(Mg,Sc)B_2$ and $Mg(B,C)_2$). In all these cases the replacing atom has the particularity of possessing one electron more outside its closed shell, $Mg{:}3s^2$ is replaced by $Al{:}3s^23p^1$ or $Sc{:}4s^23d^1$ and $B{:}2s^22p^1$ by $C{:}2s^22p^2$.

The case of *Al* replacement was studied with the *Virtual Crystal Approximation* (*VCA*) by de la Peña *et al* (2002). In this paper they found that the *Al* extra electron starts to fill up the $\sigma$-bands, and at 56% of *Al* replacement these bands become saturated and the material is no longer superconducting. They also found that the critical superconducting temperature, $T_c$, is closely proportional to the *Density of States* (*DOS*) of the $\sigma$-bands ($\sigma$-*DOS*). Kortus *et al* (2005) studied the *Al* and *C* substitutions and found that the $T_c$ drop is associated to band filling and interband scattering.

De la Mora *et al* (2005) simulated the *Al* doping in $Mg_{1-x}Al_xB_2$ by mean of the rigid band approximation (*RBA*), according to them the added charge fills up the $\sigma$-bands, and the inplane electrical conductivity of these bands, $\sigma_a^\sigma$ (the superscript runs for bands and the subscript for crystallographic direction), falls almost linearly with the Al content. This result together with the $\sigma_a^\sigma$ calculations of $MgAlB_4$ and $MgAl_2$ (50% and 100% *Al* replacement) show three parallel lines that are found as function of *Al* content. If the slope of these lines is decreased then they coincide in approximately one line. This change of slope can also be seen as a change of the doping scale, which in the frame of the RBA treatment, could be attributed to a relative shift of the $\sigma$- and $\pi$-bands due to the added *Al*-electron. This doping dependence of $\sigma_a^\sigma$, in which the three lines fall into one line shows that, with an appropriate change of scale, the *RBA* is a good description of doping. This issue will be further elaborated here.

In contrast, when *Sc* is replaced instead of *Al*, to form *(Mg,Sc)B$_2$*, the extra *3d* electron has a strong influence on the $\sigma$-bands, they become almost *3D* in *ScB$_2$* (Ruiz-Chavarria *et al* 2006) and the *RBA* can no longer be applied.

In the present paper the *Al* (*(Mg,Al)B$_2$*) and *C* (*Mg(B,C)$_2$*) doping is analyzed with the *VCA*. These results for $\sigma_a^\sigma$ as function of doping are approximately the same as those obtained with the *RBA* but with a different slope. This difference of slope in the *VCA* results could be associated to the relative shift of the $\sigma$- and $\pi$-bands. If this band-shift can be cancelled, then the *VCA* and the *RBA* results should be equivalent; changing the slope of the *VCA* results now the three lines fall approximately onto one line, especially in low doping region. Such behaviour is indicative of that, apart from the band-shift and band filling, the *Al* and *C* doping has little effect on the $\sigma$-band structure.

Detailed Band structure calculations (de la Mora *et al* 2002, and present work) show that when *MgB$_2$* is doped with *C* and *Al* the bands remain largely unchanged, the main effect is band filling, that is, $E_F$ moves upwards. There is a relative $\sigma$- and the $\pi$-band shift; for the *C*-doped system the $\sigma$-bands move down relative to the $\pi$-bands, this implies that the $\sigma$-bands fill faster. In the *Al*-doped system the shift is in the opposite direction and the $\pi$-bands have a relative down-shift, resulting in a slower $\sigma$-bands filling.

At this point it is worth mentioning that if this band-shift cancellation is applied to experimental $T_c$ data of *MgB$_2$* doped with *C* and *Al* then, an important group of experimental reports all follow the same trend, and a general $T_c$ curve for both *(Mg,Al)B$_2$* and *Mg(B,C)$_2$* systems is obtained. This general $T_c$ curve roughly follows the $\sigma$-*DOS*, the differences between the general $T_c$ curve and the $\sigma$-*DOS* can be associated to interband scattering and loss of anisotropy.

The effect of *Al* and *C* doping on the *c*-axis conductivity, $\sigma_c^\sigma$, is small, but $\sigma_c^\sigma$ is in itself small, and the effect of this doping on the $\sigma_a^\sigma/\sigma_c^\sigma$ ratio is comparatively large. This has an important impact on the anisotropy of the conductivity of the $\sigma$-bands in *(Mg,Al)B$_2$* and *Mg(B,C)$_2$* systems. From this point of view, the small difference in the $T_c$ trends of the *Al* and the *C* systems could be associated to the different loss of anisotropy in each of these systems.

## Computational details
The electronic structure calculations were done using the WIEN2k code (Blaha *et al* 2001) which is a full potential-linearized augmented plane wave (FP-LAPW) method based on DFT. The generalized gradient approximation of Perdew *et al* (1996) was used for the treatment of the exchange-correlation interactions. For the number of plane waves the used criterion was $R_{MT}^{min}$ (muffin tin radius) $\times K_{max}$ (for the plane waves) = 9. The number or *k*-points used was 19×19×15. The charge density criterion with a threshold of $10^{-4}$ was used for convergence. A denser mesh of 100×100×76 was used for the evaluation of the electrical conductivity.

**Electrical conductivity**
The electric conductivity can be calculated as in de la Mora *et al* (2005) and references within:

$$\sigma_\alpha^\beta = \frac{e^2 \tau^\beta}{\hbar \Omega_0} \int dA_\alpha \sum_i \left| v_\alpha^{i\beta}(k_F) \right| \qquad 1$$

where $\beta$ is the band index, $\tau$ is the relaxation time, $\Omega_O$ is the reciprocal-cell volume, $A_\alpha$ is the area perpendicular to the $\alpha$-direction, $v_\alpha^{i\beta}$ is the electron velocity in the $\alpha$-direction and can be calculated as the slope of the $\beta$-band ($= \hbar^{-1} \partial \varepsilon^\beta / \partial k_\alpha$), $v_\alpha^{i\beta}(k_F)$ is evaluated at $E_F$, and the sum over $i$ is for all the crossings of the $\beta$-band at $E_F$ (here $\sigma_\alpha^\beta$ refers to the conductivity, while $\sigma$ to the $\sigma$-band). From this equation the band anisotropies can be evaluated with $\sigma_a^\beta / \sigma_c^\beta$.

The band anisotropy can approximately be analyzed with the expression (de la Mora *et al* 2005):

$$\frac{\sigma_a^\beta}{\sigma_c^\beta} \approx \frac{v_a^2}{v_c^2} \approx \frac{A_a^2}{A_c^2} \qquad 2$$

The relation between conductivities, $\sigma_a^\beta / \sigma_c^\beta$, and areas, $A_a^2 / A_c^2$, was found to be correct within 17% for the *(Mg,Al)B$_2$* system (de la Mora *et al* 2005).

**Approximations**
For crystalline solid solutions in which the nuclear charge of the atoms that form the solid solution have a difference of one, for example *(Mg,Al)B$_2$* and *Mg(B,C)$_2$*, there are two simple approximations to calculate the effect of doping in the electronic structure: the Rigid Band Approximation (*RBA*) and the Virtual Crystal Approximation (*VCA*).

The *RBA* is a very simple approximation in which, as the name suggests, the band structure is kept unchanged and the Fermi energy ($E_F$) is shifted upward/downward when charge is added/subtracted. The amount of charge added/subtracted can be evaluated with the Density of States (*DOS*); it corresponds to the integration of *DOS* from $E_F$ to the shifted energy. For *(Mg,Al)B$_2$* the band structure of *MgB$_2$* is taken as initial system and $E_F$ is shifted upwards, and when a whole *e* charge is added then an approximation to the *AlB$_2$* band structure is obtained.

A better approximation is the *VCA*. When applied to *(Mg,Al)B$_2$* then the *(Mg,Al)* atoms are replaced by a virtual atom with an averaged nuclear charge; for the case of $Mg_{1-x}^{12} Al_x^{13}$ the virtual atom would have the nuclear charge of 12+x, with 0≤x≤1.

# Results and discussion
The *(Mg,Al)B$_2$* system has been studied with *VCA* by de la Peña *et al* (2002) and with *RBA* by de la Mora *et al* (2005). In this last study it was shown that the *RBA* with a suitable change of the doping scale is a good method to study the electrical conductivity on the plane, $\sigma_a^\sigma$. For this system it was also found that with electron doping, such as *Al* substitution, the $\sigma$-band conductivity decreases and the corresponding bands become less anisotropic.

In the present work the aluminium (*(Mg,Al)B$_2$*) and carbon (*Mg(B,C)$_2$*) substitutions are studied with the *VCA*. The aluminium system (*Mg$_{1-x}$Al$_x$B$_2$*) was calculated in intervals of 0.1e with $0 \leq x \leq 0.6$, while the carbon system (*MgB$_{2-x}$C$_x$*) in intervals of 0.06e with $0 \leq x \leq 0.36$. For *(Mg,Al)B$_2$* the cell parameters of de la Peña *et al* (2002) were used while for *Mg(B,C)$_2$* the cell parameters for 0, 0.06 and 0.3 were taken from the experimental values of Kazakov *et al* (2005) and then optimized with electronic structure calculations. The intermediate cell parameter values were interpolated from these values using a cubic fit. For carbon doping the usual notation is *Mg(B$_{1-x}$C$_x$)$_2$*. With the increase of *x* the extra charge added to the system is 2*x* electrons; while for the aluminium doped system, *Mg$_{1-x}$Al$_x$B$_2$*, the added charge is *x* electrons and two different scales are usually used (Gonnelli *et al* 2006, Kortus *et al* 2005), to avoid this confusion the notation $MgB_{2-x}C_x$ will be used, where *x* is the proportion of atoms replaced per unit formula which is equal to the added charge *e*.

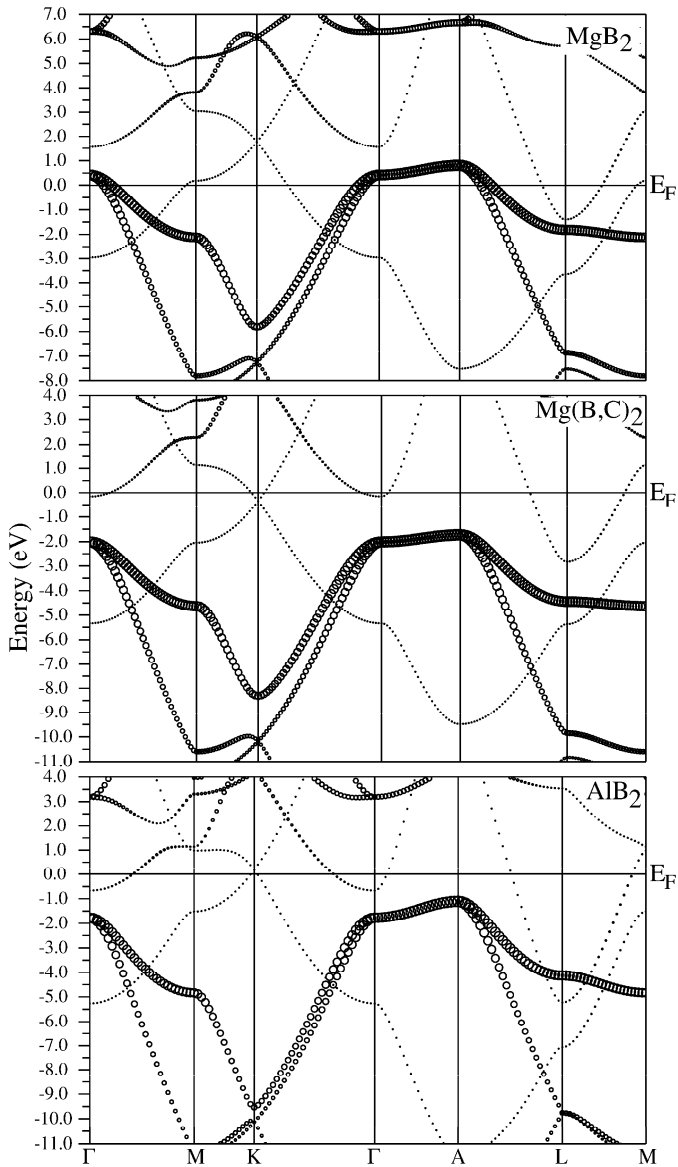

**Figure 1.** Band structure of (A) $MgB_2$, (B) $MgB_{2-x}C_x$ with *x=0.68* and (C) $AlB_2$.

Figure 1 shows the band structure calculations of a) $MgB_2$, and of b) the *C*-doped system $MgB_{2-x}C_x$, with *x=0.68* (this system was calculated using *VCA*) and of c) the *Al*-doped system $Mg_{1-x}Al_xB_2$, with *x=1.0*. It can be seen that *C*- and *Al*-doping leaves the band structure fairly unchanged; the main effect is the band-filling, that is, $E_F$ is shifted upwards. It must be pointed out that these values are quite large, since $T_c=0$ at x=0.56 in the *Al*-doped system and x=0.36 for *C* (de la Peña *et al* 2002, Kasakov *et al* 2005).

There is also a relative shift between the σ- and π-bands, in the case of *C*-doping the σ-bands move down relative to the π-bands; this implies that the band filling is faster for the σ-band. In the *Al*-doped system the relative shift is larger and in the opposite direction, therefore the σ-band fills up more slowly. Due to this difference in filling rate the *C*-doped system was chosen with *x=0.68* to account the difference of filling rate; for this doping the two systems would have comparable band structures.

In the *C*-doped system the $MgB_2$ σ-band structure is almost unchanged, while *Al*-doped system there is a small increase in the band dispersion, that is, the slope is enlarged for one of the σ-bands, this is especially noticeable at the *K* point (figure 1(B)).

An important feature in $MgB_2$ is that the σ- and π-bands are quite separated at $E_F$, this is obvious by observing the *Fermi surfaces* (*FS*) (Kortus *et al* 2001) where the σ-*FS* are tubes around the *c*-axis in the first Brillouin zone (Γ-A) while the π-*FS* are close to the external vertical sides (L-H-K-M). (Ruiz-Chavarria *et al* 2006), this feature is also valid for the *C*- and *Al*-doped systems.

The relative shift of the σ- and π-bands can be easily understood in terms of the site where the extra charge introduced by the doping is located. In the *Al*-doped system the extra charge is between the *B* planes and it affects the $p_z$-orbitals, that is, it stabilizes the π-bands relative to the σ-bands; since the π-bands are lower the added charge goes preferably to these bands and the σ-bands take longer to fill. On the other hand, for the *C* substitution the extra charge is on the *B* planes and it affects both perpendicular $p_z$-orbitals and the inplane $p_x$, $p_y$-orbitals, but the effect is stronger on the inplane orbitals;

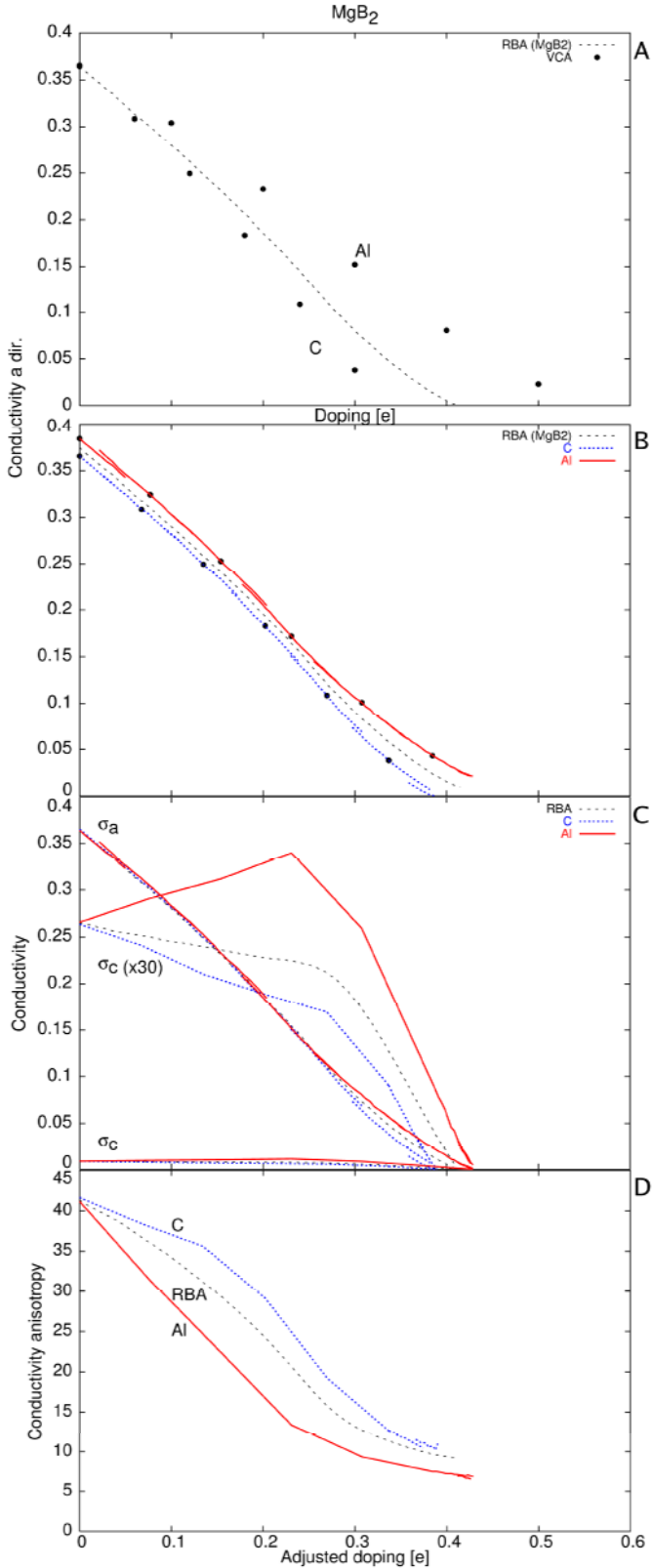

therefore $\sigma$-bands are stabilized more than the $\pi$-bands and the $\sigma$-bands fill faster. Comparing with the case of *Al*-doping, in the *C*-system the shift is in the opposite direction and smaller.

**Figure 2.** In (A) the electrical conductivity, $\sigma_a^\sigma$, is plotted as function of doping of the *C*- and *Al*-doped $MgB_2$ using *VCA*, the line refers to the *RBA* of $MgB_2$, in (B) the scale is adjusted by multiplying by a constant (1.12 for *C*-doped, and 0.77 for *Al*-doped), the plots are shifted upwards by 0.02 for visibility and the *VCA* results are interpolated with *RBA*, in (C) the inplane conductivity, $\sigma_a^\sigma$, and the *c*-direction conductivity, $\sigma_c^\sigma$, are plotted as function of adjusted doping; $\sigma_c^\sigma$ is also shown multiplied ×30 for visibility, in (D) the anisotropy of the conductivity, $\sigma_a^\sigma/\sigma_c^\sigma$, is plotted as function of adjusted doping.

If the only effect of doping were an upward shift of $E_F$ then, for properties that depend on $E_F$ only, such as the electrical conductivity, and the *RBA* would be a very good description. In the case of $MgB_2$ the superconductivity seems to be dependent especially on the properties of the $\sigma$-bands at $E_F$ (de la Peña *et al* 2002, de la Mora *et al* 2005, Canfield and Crabtree 2003). This suggests that for properties related to the $\sigma$-band at $E_F$, the relative $\sigma$- and $\pi$-band shift, which speeds or slows, the $\sigma$-band filling could be taken into account by a change of doping scale.

The *VCA* results for the electrical conductivity on the *a-b* plane, $\sigma_a^\sigma$, for *C* and *Al* doping in $MgB_2$ are shown as dots in figure 2(A), the *RBA* on $MgB_2$ (*RBA-MgB₂*) is the continuous line in between. Following the ideas just mentioned, by changing the doping scale, these *VCA* results could be made equivalent to the *RBA-MgB₂*. This effect is achieved by multiplying by 1.12 the scale for the *C*-doped system and by 0.77 the *Al*-doped system (see figure 2(B); at zero doping the three results are the same, but for visibility the results were shifted upward by 0.01). These *VCA* results can be interpolated using *RBA* and an almost continuous line for both the *C*- and *Al*-doped systems is found. The unshifted curves can be seen as $\sigma_a$ in figure 2(C). It can be seen that the *VCA* and the *RBA* curves are almost the same. It is only at high doping values that the results start to diverge.

This coincidence of the *VCA* (adjusted with a change of doping scale) with the *RBA-MgB₂* shows that for the $\sigma_a^\sigma$ the main effect of the *C* and *Al* substitutions in $MgB_2$ is associated to the band-filling and band-shift. Contrary to these cases is the effect of *Sc* substitution, *(Mg,Sc)B$_2$*, in which the *Sc:3d* orbital strongly affects the $\sigma$-bands and the *RBA-MgB$_2$* with a change of scale cannot be used to study this system (Ruiz-Chavarria *et al* 2006).

The band structure of the doped systems (figure 1(B) and (C)) were calculated with respect to *MgB$_2$* (figure 1(A)) at doping values about 90% higher than that point where the conductivity becomes zero, that is, when the $\sigma$-bands become saturated. This high value was chosen so that the effects of doping on the band structure become clear.

The conductivity in the *c*-direction, $\sigma_c^\sigma$, of both *Al*- and *C-MgB$_2$* doped systems is very small and is shown in the lower part of figure 2(C), for visibility it is also shown multiplied ×30. For the case of *Al* substitution, $\sigma_c^\sigma$ first increases then drops; in the case of *C* substitution, $\sigma_c^\sigma$ drops continuously. The *RBA* gives a result in between the two *Al* and *C* substitutions. This $\sigma_c^\sigma$ conductivity is substantially modified by the *VCA* in relation to the *RBA*; for the *Al* system it increases and for the *C* system it diminishes. This could not be justified simply by a change of cell parameters since in both cases the *c*-parameter diminishes (de la Peña *et al* 2002, Kazakov *et al* 2005)

It is interesting to notice that the $\sigma_c^\sigma$ term cannot be interpolated with *RBA* below the *Electronic Topological Transition* (*ETT*) (Bianconi *et al* 2002 and Agrestini *et al* 2004), which corresponds to the discontinuity point where the $\sigma$-band *FS* change from tubes to ellipsoids (≈ 0.25e). This situation can be intuitively understood in terms of expression 2; before the *ETT* these *FS* are open tubes and the area seen from above is a ring. With increasing doping the inner circle closes, in this case the area of the ring increases relatively fast and at the *ETT* this inner circle disappears and the ring becomes a full circle and the area changes at a lower pace with doping.

The anisotropy of the bands can be associated with the anisotropy of the electrical conductivity, $\sigma_a^\sigma/\sigma_c^\sigma$. This anisotropy drops linearly for the *Al* substitution and then levels off at high doping, see figure 2(C). For the *C* substitution the behaviour is similar but it starts to drop more slowly. At the highest doping level the anisotropy for both cases reaches the low value of ~10 and these systems cannot be regarded as *2D* any more. It should be pointed out that the final anisotropy for both systems is of the same order as the anisotropy estimated for the *(Mg,Sc)B$_2$* system, in which the loss of $T_c$ cannot be associated only to the loss of $\sigma$-carriers but it should also be associated to the loss of anisotropy. In this system the loss of anisotropy seems to play an essential role (Ruiz-Chavarria *et al* 2006). If in the *Sc*-doped system the anisotropy is an important factor, then in the *C*- and *Al*- doped systems this anisotropy should also play an important role.

Does the relative $\sigma$- and $\pi$-band-shifts, which give different doping scales to the *C* and *Al* systems, affect other properties in *Mg(B,C)$_2$* and *(Mg,Al)B$_2$*? We think that it does. The respective drop of the critical temperatures, $T_c$, follow different curves, but, as we will show, by applying the change of doping scale, as was done for the conductivity, reduces the two $T_c$ curves into one.

The experimental $T_c$ for *C* and *Al* doping are plotted in figure 3(A) (Bianconi *et al* 2002, Postorino *et al* 2001, Slusky *et al* 2001 and Putti *et al* 2003 for the *(Mg,Al)B$_2$* system and Kasakov *et al* 2005, Lee *et al* 2003 and Bharathi *et al* 2002 for the *Mg(C,B)$_2$* system and Gonnelli *et al* 2006 for both systems). Only $T_c$ experimental papers with a large doping range were taken into account. The lines in that graph correspond to simple cubic fits on each data set.

As mentioned above, the change of doping scale corresponds to an increment of the doping axis by 1.12 in the case of the *C*-system, and a reduction by 0.77 in the case of the *Al*-system. The $T_c$ plots with adjusted doping are shown in figure 3(B); here a dramatic reduction of dispersion in $T_c$ is observed.

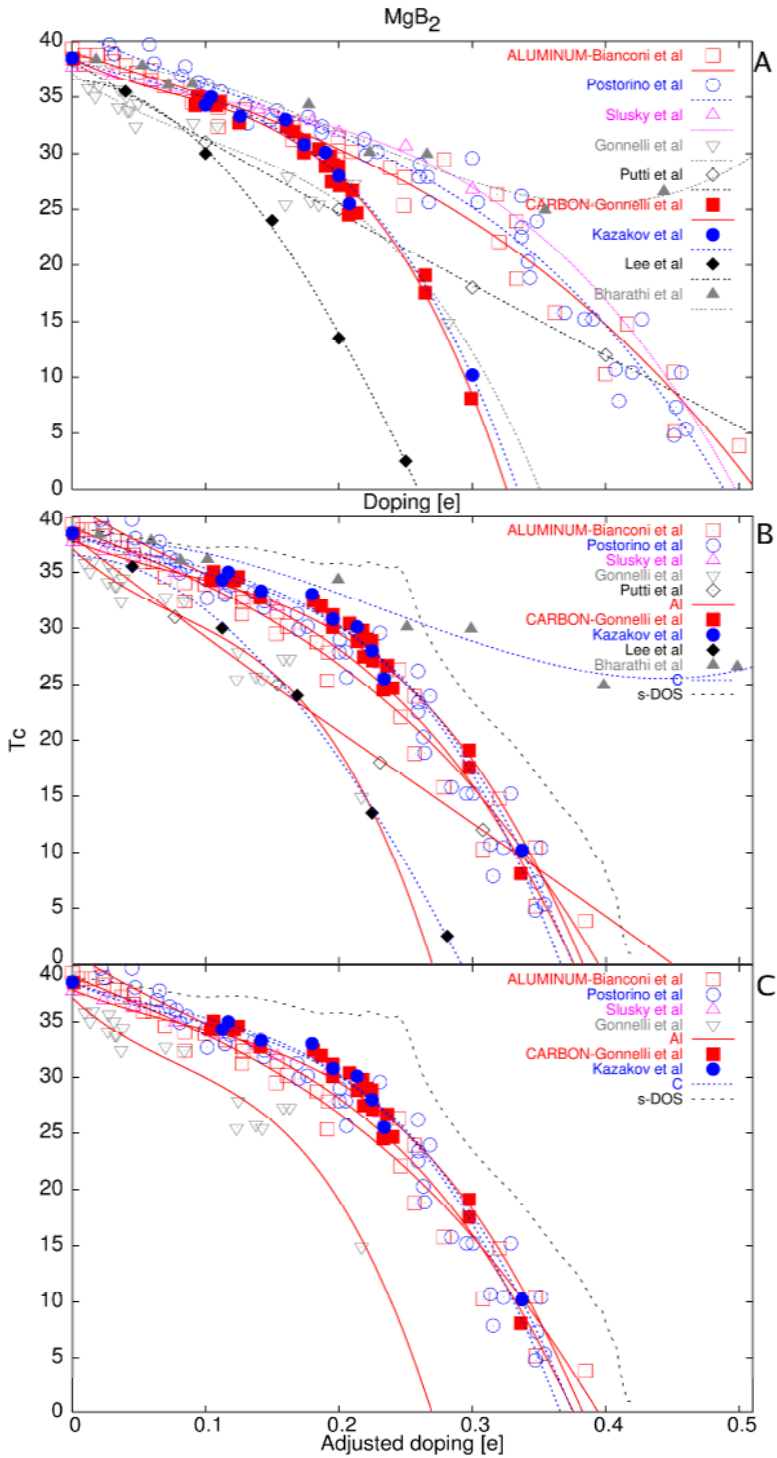

**Figure 3.** (A) Experimental $T_c$ data of *C*- and *Al*-doped $MgB_2$ as function of doping, in (B) they are plotted as function of adjusted doping, (C) selected experimental $T_c$ data as function of adjusted doping.

Most of the data sets show a general downward turn, with $T_c$ tending to zero at ~0.36 *e* for the *C*-doping, and at ~0.5 *e* for the *Al*-doping. The data of Putti *et al* (2003) (*Al*-doping) does not have the downward turn, this could be because that they used the nominal concentrations, this is also the case of the data of Lee *et al* (2003) (*C*-doping). The data of Bharathi *et al* (2002), for *C*-doping, follows a pattern very different from all the other sets, it extends up to 0.44 *e* doping but maintaining relatively high $T_c$ values.

The rest of the experimental $T_c$ data-sets are shown in figure 3(C) (Bianconi *et al* 2002, Postorino *et al* 2001 and Slusky *et al* 2001 for the $(Mg,Al)B_2$ system and Kasakov *et al* 2005 in the $Mg(C,B)_2$ system and Gonnelli *et al* 2006 for both systems). The *Al*-data of Gonnelli *et al* has only one point at high doping and differs noticeably from the other sets.

The dotted line in this figure is the $p_\sigma$ ($p_x + p_y$) contribution to *DOS* ($p_\sigma$-*DOS*), which is quite similar to the $\sigma$-band *DOS* (Mazin and Antropov 2003, Ann *et al* 2001). The $p_\sigma$-*DOS* reflects the almost *2D*-character of the $\sigma$-*FS*, it is fairly constant at low doping, and at the *ETT* has a sharp downturn and then it diminishes relatively fast. The data sets without that of Gonnelli *et al* (2006) (for *Al*) all follow a very similar trend, suggesting a universal trend. On the other hand, in this work they present data for both *C* and *Al*; one of the co-workers, J. Karpinski (2006), provided us with new data for both systems that corroborates the trends for both *C* and *Al* systems.

This universal trend of the experimental $T_c$ data-sets was achieved by compensating for the relative $\sigma$- and $\pi$-band-shifts. The meaning of such compensation is that, if there were no relative band shifts then, both the inplane $\sigma$-conductivity, $\sigma_a^\sigma$, and the $T_c$ would follow the same general trend in the two systems. This, we think, is an important result since it shows that with the appropriately adjusted doping-axis, the *RBA* represents a good description of the electronic properties of these systems. In other words, the main effect of the extra *p*-electron of the *Al* and *C* systems is to fill the $\sigma$-bands; the second effect is the relative $\sigma$- and $\pi$-band-shift. These results contrasts with that of the *Sc*-doped system *((Mg,Sc)B$_2$)* in which the bands are substantially modified by the doping (Ruiz-Chavarria *et al* 2006). The *Sc* extra *d*-electron significantly affects the $\sigma$-bands; these are not saturated even in the fully doped system *ScB$_2$*, and $\sigma_a^\sigma$ diminishes slowly, but does not disappear; on the other hand, the anisotropy of *ScB$_2$*, $\sigma_a^\sigma/\sigma_c^\sigma$, quickly diminishes at relatively low doping (0-30%).

The experimental $T_c$ data sets approximately follow the $p_\sigma$-*DOS* curve. Even the sharp downturn at *ETT* in the $p_\sigma$-*DOS* is faithfully reflected in the *Al* data sets of Bianconi *et al* (2002), Postorino *et al* (2001) and Slusky *et al* (2001), where a small change of slope can be observed at ~0.25 *e* of the adjusted doping. On the other hand, there two are important differences that can be noticed: at low doping $p_\sigma$-*DOS* is almost constant, while $T_c$ drops faster, as well the $T_c$ drop to zero occurs at lower doping value.

Bianconi *et al* (2002) associate a *2D→3D* transition to the *ETT*. $T_c$ shows a small change of slope at the *ETT*, but the electrical anisotropy, $\sigma_a^\sigma/\sigma_c^\sigma$, does not show any change at this value; instead there is a smooth dimensional transition that spans a large doping range. For the *(Mg,Sc)B$_2$* system, Agrestini *et al* (2004) also associate the dimensional transition to the *ETT*, and again the electrical anisotropy shows a smooth dimensional transition spanning a large doping range (Ruiz-Chavarria *et al* 2006).

The similarity between the *Al* experimental $T_c$ data set of Bianconi *et al* (2002) and the $\sigma$-*DOS* curve was observed earlier by de la Peña *et al* (2002); they found an almost perfect fit between $T_c$ and the *FS* area which is closely related to the $\sigma$-*DOS*. Kortus *et al* (2005) analyzed the *C* and *Al* systems and found that the $T_c$ drop is due to band filling and interband scattering, although they were not able to find a general trend for $T_c$.

The relative constancy of the $p_\sigma$-*DOS* at low doping in *(Mg,Al)B$_2$* and *Mg(C,B)$_2$* would imply a very small band-filling; therefore if the $T_c$ drop is due to this band-filling, it should remain fairly constant, what is observed is a fairly large decrease. This difference could be associated to interband scattering (Kortus *et al* 2005) and to loss of anisotropy of the systems (Ruiz-Chavarria *et al* 2006). In doped *MgB$_2$* there should be little interband scattering, since the $\sigma$- and $\pi$-bands are quite separated, and as was commented by Andersen (2006) they 'do not talk to each other', the same was found by Mazin *et al* (2002) for low quality samples of *MgB$_2$*. Therefore the loss of anisotropy seems to be the most probable cause of this $T_c$ drop.

Furthermore, interband scattering, as Kortus *et al* argue, would have a larger effect on the *C*-doped system due to the fact that *C* is a stronger scatterer, since it contributes to both the $\sigma$- and $\pi$-bands, while *Al* contributes only to the $\pi$-bands. Therefore $T_c$ is expected to drop faster for these *C*-doped systems. What is experimentally observed is the opposite, the $T_c$ curves for the *C*-doped systems are slightly above than those for the *Al*-doped systems; this is not taking into account the *Al* data of Gonnelli *et al* (2006) which are clearly below all the curves, and would increase this difference even further. Therefore, if interband scattering is present, it is playing a smaller role than anisotropy loss.

**Conclusions**
Band structure calculations using the virtual crystal approximation show that the main effect of *C* and *Al* doping on *MgB$_2$* is band filling; the $\sigma$-bands are little deformed by this doping. A smaller effect is a relative shift between the $\sigma$- and $\pi$-bands. For features that depend only on the $\sigma$-bands at $E_F$, this shift can be taken into account or eliminated by an appropriate change of doping scale.

Applying this change of scale to the inplane $\sigma$-band electrical-conductivity, the resulting curves almost coincide with the case where there is no band-shift and the $\sigma$-bands are not deformed; that is, the rigid band approximation. The superconducting critical temperature of the *C*- and *Al*-doped systems with this change of scale now follow almost the same curve, which approximately follows the $\sigma$-band density of states. The difference between the critical temperature and this density of states can be associated to interband scattering and loss of anisotropy, although it has been found that interband scattering is small in these systems and also the loss of anisotropy is comparable of that observed in the scandium doped *MgB$_2$*, in which at the highest doping value the $\sigma$-bands cannot longer be regarded as two-dimensional.

The conductivity in the *c*-direction is small, in this case doping has a relative large effect. Therefore the anisotropy drops faster for the *C*-doped systems; this is reflected in the critical temperature curves. Therefore, the drop of anisotropy and the differences of anisotropy in the *C*- and *Al*-systems and their relation with the experimental $T_c$ curves show that the loss of anisotropy is playing an important role in these systems.

## Acknowledgements

We thank Jens Karpinsky for valuable discussions and for sending their experimental data, also to Romeo de Coss for his useful comments. This work was done with support from DGAPA-UNAM under project PAPIIT112005